\documentclass[11pt,a4paper,thmsa,final,ukenglish]{article}
\usepackage[T1]{fontenc}
\usepackage[latin9]{inputenc}
\usepackage{amsmath}
\usepackage{graphicx}

\makeatletter

\pdfpageheight\paperheight
\pdfpagewidth\paperwidth

\newcommand{\lyxmathsym}[1]{\ifmmode\begingroup\def\b@ld{bold}
  \text{\ifx\math@version\b@ld\bfseries\fi#1}\endgroup\else#1\fi}

\providecommand{\tabularnewline}{\\}

\usepackage{textcomp}
\makeatother

\begin{document}

\title{A preliminary DFT study of the adsorption and dissociation of$CH_{4}$,
$SO_{2}$ and $O_{2}$ reactions on $Cr_{2}O_{3}(0001)$}

\author{S. N. Hernández Guiance\textsuperscript{a},
D. Coria\textsuperscript{b}, I. M. Irurzun\textsuperscript{{*},c},
E. E. Mola\textsuperscript{1}.}

\maketitle
\textsuperscript{a}Facultad de Química, e Ingeniería, Pontificia
Universidad Católica Argentina, Mendoza 4197 CP 2000 Rosario, Argentina.

\textsuperscript{b}Universidad del Centro Educativo Latinoamericano.
Av. Pellegrini 1332 CP 2000 Rosario, Argentina.

\textsuperscript{c}CCT La Plata \textendash{} CONICET. Instituto
de Investigaciones Fisicoquímicas Teóricas y Aplicadas (INIFTA), Facultad
de Ciencias Exactas, Universidad Nacional de La Plata. Diagonal 113
y 64, CP (1900), La Plata, Argentina. Telephone: +54 221 425 74 30,
ext. 149. Fax: +54 221 425 46 42.

{*}Correspondig Author

\textsuperscript{1} In memory 
\begin{abstract}
In the present work, we study the structures and molecular geometries
of $CH_{4}$, $SO_{2}$ and $O_{2}$ adsorbed on $Cr_{2}O_{3}(0001)$.
Using computational calculations based on the density functional theory
(DFT), we analyze the most suitable sites to carry out the adsorption
of each of the molecules mentioned, and the influence of each species
on the adsorption and dissociation of the others. The results allow
us to understand the activation of the $Cr_{2}O_{3}(0001)$ surface,
which leads to the presence of $SO_{2}$ during the oxidation of $CH_{4}$,
as was experimentally verified. 
\end{abstract}

\section{Introduction}

Transition metal oxide surfaces have been a major focus of interest
in the field of catalysis and corrosion. Environmental sciences have
paid much attention to catalytic surfaces, which have been used to
remove pollutant molecules, such as $SO_{2}$, $CH_{4}$, $CO$, and
$CO_{2}$, from the atmosphere. Among these molecules, $SO_{2}$ is
one of the products from natural and anthropogenic sources emitted
to the atmosphere that may turn into acid rain. Most of the $SO_{2}$
released into the atmosphere (\ensuremath{\approx} 3/4) is produced
by human activities, especially by the combustion of fossil fuels.
More than half of the world production comes from a few developed
countries. In industrial stacks, $SO_{2}$ reduction occurs by reaction
with $CH_{4}$, which is then oxidized to $CO_{2}$ by $O_{2}$. Therefore,
the following reactions

\begin{equation}
CH_{4}+2SO_{2}\xrightarrow{k_{1}}2S^{0}+CO_{2}+2H_{2}O
\end{equation}

\begin{equation}
2O_{2}+CH_{4}\xrightarrow{k_{2}}k_{2}CO_{2}+2H_{2}O
\end{equation}

\begin{equation}
SO_{2}+CH_{4}+O_{2}\xrightarrow{k_{3}}S^{0}+CO_{2}+2H_{2}O
\end{equation}

as well as the development of catalysts that favor $CH_{4}$ oxidation
in the presence of $SO_{2}$ and $O_{2}$ have been a major subject
of research. In previous work, we studied the oxides of some transition
metals ($Co$, $Ni$, $Fe$, $V$, $Mn$, $Cr$, and $Mo$) supported
on alumina \cite{key-1} in order to determine the best surface for
capturing $SO_{2}$. We found that the best adsorbent is $Cr_{2}O_{3}/Al_{2}O_{3}$
\cite{key-2,key-3}, which also catalyzes $CH_{4}$ oxidation to obtain
$CO_{2}$. It was selected due to its thermal and mechanical resistance
and its possibility to be regenerated. We also experimentally determined
the activation energies of the aforementioned reactions on $Cr_{2}O_{3}/Al_{2}O_{3}$
under stoichiometric conditions \cite{key-4}. We concluded that the
presence of $SO_{2}$ favors $CH_{4}$ oxidation, decreasing the activation
energy. In this paper, we present theoretical results based on the
density functional theory on the adsorption and dissociation of diverse
species involved in reactions (1)\textendash (3) on $Cr_{2}O_{3}/Al_{2}O_{3}$.
We aim to provide additional theoretical information to understand
the mechanism of the oxidation reaction of $CH_{4}$ on$Cr_{2}O_{3}/Al_{2}O_{3}$
in the presence of $SO_{2}$ and $O_{2}$.

\section{Theoretical}

First-principles total energy calculations were performed using DFT+U
to investigate individual and simultaneous adsorption of $SO_{2}$,
$CH_{4}$ and $O_{2}$ molecules on the $\alpha-Cr_{2}O_{3}(0001)$
surface as implemented in the Vienna Ab initio Simulation Package
(VASP) code.

The $\alpha-Cr_{2}O_{3}(0001)$ surface has been studied using different
theoretical approaches \cite{key-6,key-7,key-8,key-9,key-10,key-11}.
There is general agreement among the different methods that the surface
undergoes strong vertical relaxations. In this paper, we take into
account the strong correlation effects described by a Hubbard-type
on-site Coulomb repulsion, not included in a density functional description
\cite{key-10}. The $(0001)$ face was selected because in its natural
state, $Cr_{2}O_{3}$ has this type of structure in $97.20\%$ of
its volume, which is maintained up to temperatures of about $973K$.

The Kohn-Sham equations were solved using projector augmented wave
(PAW) method and a plane-wave basis set including plane waves up to
400 eV. Electron exchange and correlation energies were calculated
within the local spin density approximation (LSDA) in the Perdew-Zunger
form. The DFT+U method was used with values J=1 and U=5 \cite{key-10}.
Convergence is considered achievedwhen the forces on the ions are
less than 0.03 eV/Å. Periodic boundary conditions are applied in the
three perpendicular directions. The Hessianmatrix of second derivativeswas
determined for ground structures within the harmonic approximation
by two-sided finite differences, using a displacement step of 0.01
Å. Adsorbed atoms were displaced in the calculations, and diagonalization
of the dynamic matrix yields the harmonic frequencies.

The surface is modelled as a rhomboid supercell with an edge size
of 4.954Å and 20Å high. Each substrate layer is composed of one chromium
atom, three oxygen atoms and one chromium atom, and is 2.263 Å thick.
However, hereafter each layer composed of one type of atom only will
be called a layer. The supercell used is shown in Fig. 1, while the
geometric parameters obtained after surface optimization are listed
in Table 1.

\begin{table}
\begin{tabular}{|c|c|c|}
\hline 
Interlayer spacing  & DFT+U LSD  & GGA\tabularnewline
\hline 
\hline 
$Cr_{1}-O_{2}$  & -61  & -53\tabularnewline
\hline 
$O_{2}-Cr_{3}$  & +6  & +14\tabularnewline
\hline 
$Cr_{3}-Cr_{4}$  & -44  & +70\tabularnewline
\hline 
$Cr_{4}-O_{5}$  & +9  & +12\tabularnewline
\hline 
$O_{5}-Cr_{6}$  & -2  & +12\tabularnewline
\hline 
$Cr_{6}-Cr_{7}$  & +7  & -56\tabularnewline
\hline 
$Cr_{7}-O_{8}$  & -2  & +10\tabularnewline
\hline 
$O_{8}-Cr_{9}$  & -1  & -5\tabularnewline
\hline 
$Cr_{9}-Cr_{10}$  & +3  & -5\tabularnewline
\hline 
\end{tabular}

\caption{Variation in atomic interlayer spacings of the $\alpha-Cr2O3(0001)$
substrate after surface optimization. The initial spacings are: $Cr-O=0.94\mathring{A}$,
$Cr-Cr=0.38\mathring{A}$. Comparison of the results from DFT, LSDA
and GGA methods.}
\end{table}

\begin{figure}[h]
\begin{centering}
\includegraphics[height=4cm]{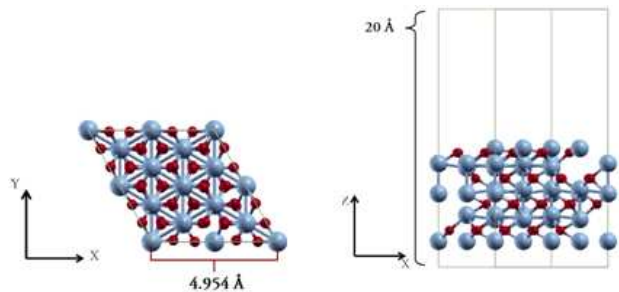} 
\par\end{centering}
\caption{$Cr_{2}O_{3}(0001)$ supercell.}
\end{figure}

The first Brillouin zone was sampled with a ($3\times3\times1$) gamma
point centered mesh and only the gamma point was used for the cubic
supercell used for the optimization of isolated molecules \cite{key-3}.

The adsorption energy of each adsorbate molecule is calculated as

\begin{equation}
E_{a}=E(Adsorbate/Cr_{2}O_{3})\lyxmathsym{\textendash}E(Adsorbate)\lyxmathsym{\textendash}E(Cr_{2}O_{3})
\end{equation}

The first term is the energy of the optimized configuration for the
relaxed adsorbate molecule bonded to the clean surface. The second
term is the energy of the optimized gas-phase adsorbate molecule (isolated),
and the third term is the energy of the optimized surface. Based on
this definition, negative values of $E_{a}$ correspond to stable
configurations. The structures of each isolated molecule were optimized
first; they are called ``simples systems\textquotedblright :

- $SO_{2}$on $Cr_{2}O_{3}(0001)$

- $CH_{4}$on $Cr_{2}O_{3}(0001)$

- $O_{2}$on $Cr_{2}O_{3}(0001)$

- $S$ on $Cr_{2}O_{3}(0001)$

- $CO$ on $Cr_{2}O_{3}(0001)$

- $CO_{2}$on $Cr_{2}O_{3}(0001)$

Using the results from the most stable geometries of these systems,
molecules of other chemical species were adsorbed; they are called
``compound systems\textquotedblright :

- $SO_{2}$ on $O_{2}$ preadsorbed on $Cr_{2}O_{3}(0001)$.

- $CH_{4}$ on $O_{2}$ preadsorbed on $Cr_{2}O_{3}(0001)$.

- $O_{2}$ on $SO_{2}$, preadsorbed on $Cr_{2}O_{3}(0001)$.

In a first stage, the calculations were performed using pseudopotentials
based on the local density approximation (LDA) \cite{key-5,key-6}.
Then, in order to achieve greater accuracy, pseudopotentials within
the generalized gradient approximation (GGA) were used.

\section{Results}

\subsection{Simple systems: adsorbate on substrate}

- $SO_{2}$ on $Cr_{2}O_{3}(0001)$: after generating a clean surface
and optimizing each isolated molecule, a $SO_{2}$ molecule was adsorbed
on the $Cr_{2}O_{3}(0001)$ surface in different positions and geometries.
The most stable configuration was found in previous studies \cite{key-2}
using the LDA pseudopotentials. The final results are listed in Table
2.

\begin{table}
\begin{tabular}{|c|c|c|c|c|}
\hline 
$E_{a}$(eV)  & X-Y Plane  & X-Z Plane  & $D\left[S-O_{ads}\right](\mathring{A})$  & $D\left[S-Cr\right](\mathring{A})$\tabularnewline
\hline 
\hline 
-3.09  & \includegraphics[height=2.5cm]{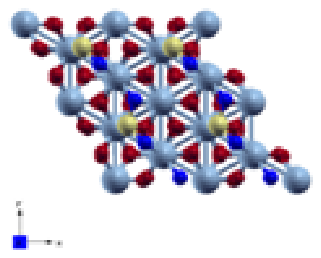}  & \includegraphics[height=2.5cm]{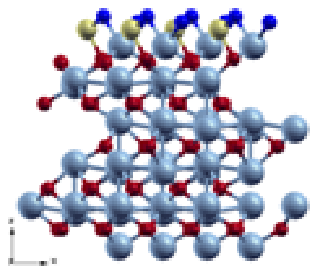}  & 1.51  & 1.77\tabularnewline
\hline 
\end{tabular}

\caption{Most stable geometry of the $SO_{2}$ system on $Cr_{2}O_{3}$ optimized
with the DFT+U formalism and its activation energy.}
\end{table}

The adsorption of $SO_{2}$ on the $Cr_{2}O_{3}$ surface is a chemisorption
process involving sulfite species formation and surface oxygen atoms:
the sulfur atom binds to one surface oxygen atom, while sulfur dioxide
oxygen atoms are bound to chromium atoms.

- $CH_{4}$ on $Cr_{2}O_{3}(0001)$: one $CH_{4}$ molecule was adsorbed
on the $Cr_{2}O_{3}(0001)$ surface in different positions and geometries.
The most stable results, which are summarized in Table 3, indicate
that no molecular adsorption occurred on the surface under study.

\begin{table}
\begin{tabular}{|c|c|c|c|c|c|}
\hline 
$E_{a}$(eV)  & X-Y Plane  & X-Z Plane  & $D\left[C-Cr_{sup}\right](\mathring{A})$  & $D\left[H_{1}-Cr_{sup}\right](\mathring{A})$  & $D\left[H_{2}-Cr_{sup}\right](\mathring{A})$\tabularnewline
\hline 
\hline 
-0.0159  & \includegraphics[height=2.5cm]{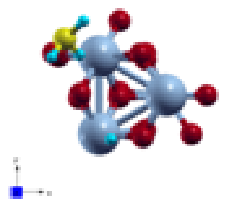}  & \includegraphics[height=2.5cm]{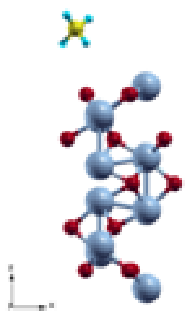}  & 5.255  & 5.836  & 5.438\tabularnewline
\hline 
-0.0138  & \includegraphics[height=2.5cm]{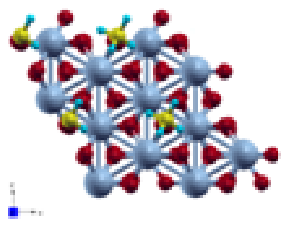}  & \includegraphics[height=2.5cm]{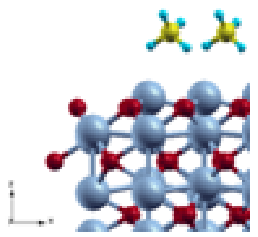}  & 5.235  & 5.532  & 4.155\tabularnewline
\hline 
-0.014  & \includegraphics[height=2.5cm]{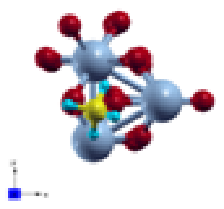}  & \includegraphics[height=2.5cm]{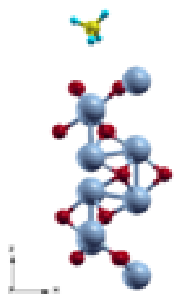}  & 3.783  & 4.752  & 4.124\tabularnewline
\hline 
\end{tabular}

\caption{Most stable geometries of the $CH_{4}$ system on $Cr_{2}O_{3}$ optimized
with the DFT+U formalism and their activation energies.}
\end{table}

- $O_{2}$ on $Cr_{2}O_{3}(0001)$: the most stable results for this
system are shown in Table 4 and Fig. 2. The two atoms of the oxygen
molecules are adsorbed on the same surface chromium atom.

\begin{table}
\begin{tabular}{|c|c|c|c|c|c|}
\hline 
$E_{a}$(eV)  & X-Y Plane  & X-Z Plane  & $D\left[O_{1}-O_{2}\right](\mathring{A})$  & $D\left[O_{1}-Cr_{3}\right](\mathring{A})$  & $D\left[O_{2}-Cr_{2}\right](\mathring{A})$\tabularnewline
\hline 
\hline 
-0.478  & \includegraphics[height=2.5cm]{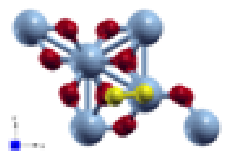}  & \includegraphics[height=2.5cm]{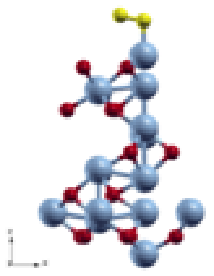}  & 1.214  & 3.054  & 2.136\tabularnewline
\hline 
-0.446  & \includegraphics[height=2.5cm]{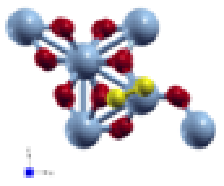}  & \includegraphics[height=2.5cm]{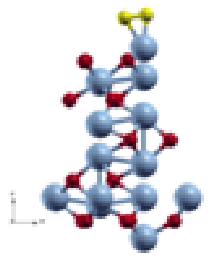}  & 1.344  & 2.002  & 1.993\tabularnewline
\end{tabular}

\caption{Most stable geometries of the $O_{2}$ molecularly adsorbed on $Cr_{2}O_{3}$
optimized with the DFT+U formalism and their activation energies.}
\end{table}

\begin{figure}
\begin{centering}
\includegraphics[width=12cm]{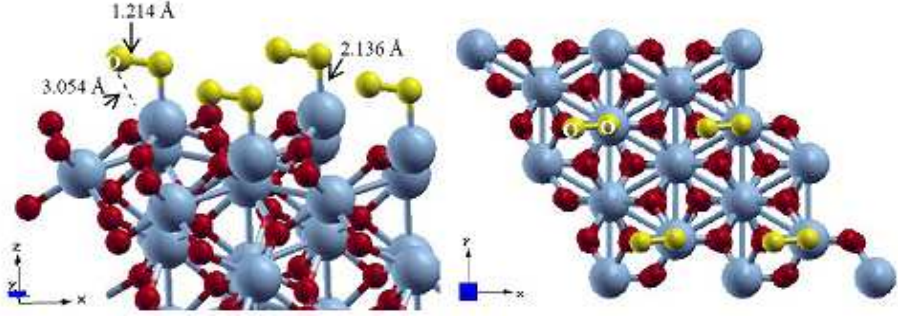} 
\par\end{centering}
\caption{Geometric details of the first structure listed in Table 4.}
\end{figure}

- Dissociative $O_{2}$ adsorption on $Cr_{2}O_{3}(0001)$: in this
system, two $O_{2}$ atoms were placed on the surface in order to
study the stability of the adsorbate molecule and the possible availability
of oxygen atoms to oxidize methane to $CO_{2}$. The most stable dissociative
adsorption configuration shows both $O_{2}$ atoms adsorbed on the
same chromium atom in the first layer (see Table 5 and Fig. 3).

\begin{table}
\begin{tabular}{|c|c|c|c|c|c|}
\hline 
$E_{a}$(eV)  & X-Y Plane  & X-Z Plane  & $D\left[O_{1}-O_{2}\right](\mathring{A})$  & $D\left[O_{1}-Cr_{3}\right](\mathring{A})$  & $D\left[O_{2}-Cr_{2}\right](\mathring{A})$\tabularnewline
\hline 
\hline 
-0.474  & \includegraphics[height=2.5cm]{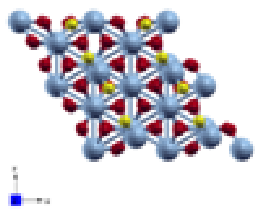}  & \includegraphics[height=2.5cm]{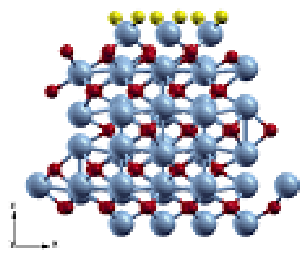}  & 2.466  & 1.585  & 3.85\tabularnewline
\end{tabular}

\caption{Most stable geometry of the dissociative oxygen adsorption on $Cr_{2}O_{3}$
optimized with the DFT+U formalism and its activation energy.}
\end{table}

\begin{figure}
\begin{centering}
\includegraphics[width=12cm]{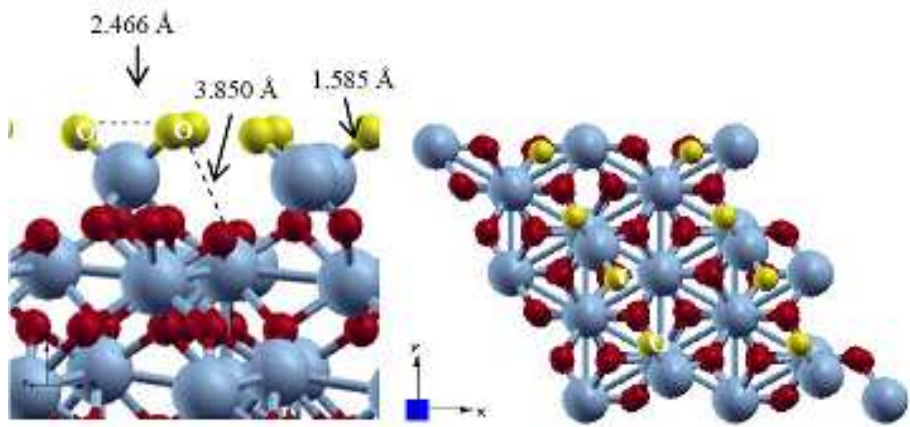} 
\par\end{centering}
\caption{Geometric details of the first structure listed in Table 5.}
\end{figure}

- $S$ on $Cr_{2}O_{3}(0001)$: a sulfur atom was adsorbed on the
$Cr_{2}O_{3}(0001)$ surface at five sites: on one chromium atom of
the first layer, obtaining the lowest adsorption energy of -0.594
eV; on a chromium atom of the third layer; and on an oxygen atom of
the second, fourth and fifth layers \cite{key-2,key-5}.

\begin{figure}
\begin{centering}
\includegraphics[height=4cm]{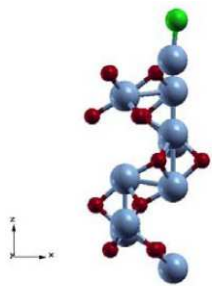} 
\par\end{centering}
\caption{Most stable optimized structure of $S$ adsorbed on the $Cr_{2}O_{3}(0001)$
surface.}
\end{figure}

- $CO$ on $Cr_{2}O_{3}(0001)$: one $CO$ molecule was adsorbed on
the $Cr_{2}O_{3}(0001)$ surface, obtaining an adsorption energy of
-2.33 eV in the most stable geometry {[}5{]}.

- $CO_{2}$ on $Cr_{2}O_{3}(0001)$: one $CO_{2}$ molecule was adsorbed
in different positions and geometries. The system obtained was more
unstable than that of $CO$ on the same surface, with an adsorption
energy of -0.81 eV for the most stable geometry (see Fig. 5).

\begin{figure}
\begin{centering}
\includegraphics[height=4cm]{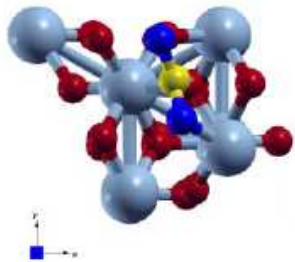} 
\par\end{centering}
\caption{Most stable optimized geometry of $CO_{2}$ adsorbed on the $Cr_{2}O_{3}(0001)$
surface.}
\end{figure}

\subsection{Complex systems: adsorption of one species on another one preadsorbed
on the substrate}

After obtaining the most stable geometries for the adsorption of one
molecule on the catalytic surface, we investigated the adsorption
of one of the species under study on another one preadsorbed on $Cr_{2}O_{3}(0001)$.
Thus, the interaction between gas species on the surface was examined,
obtaining the adsorption energies of the most stable configuration,
together with the bond lengths, the variations in the angles formed
between the atoms of those species, and the new compounds formed once
the interactions between them had ceased.

- $CH_{4}$ on an $O_{2}$ molecule preadsorbed on $Cr_{2}O_{3}(0001)$:
the adsorption of one methane molecule on the $Cr_{2}O_{3}(0001)$
surface was studied using the most stable geometry for the adsorption
of one $O_{2}$ molecule. No stable geometry was found, indicating
that methane does not adsorb on the $O_{2}$ molecule adsorbed on
$Cr_{2}O_{3}(0001)$.

- $CH_{4}$ on $O_{2}$ dissociatively adsorbed on $Cr_{2}O_{3}(0001)$:
a $CH_{4}$ molecule was placed in different positions on an $O_{2}$
molecule dissociatively adsorbed on the substrate. In this case, as
in the above, the most stable geometry for the dissociative $O_{2}$
adsorption on $Cr_{2}O_{3}(0001)$ was used. The dissociation of the
methane molecule was obtained, as well as the formation of $OH$ species
and methoxy groups ($-O-CH_{3}$) as shown in Fig. 6.

\begin{figure}
\begin{centering}
\includegraphics[width=12cm]{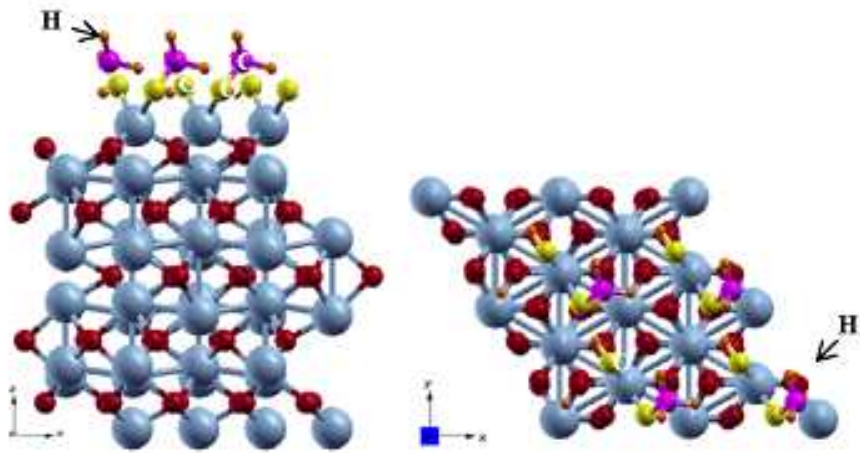} 
\par\end{centering}
\caption{Most stable optimized geometry of $CH_{4}$ on $O_{2}$ dissociatively
adsorbed on the $Cr_{2}O_{3}(0001)$ surface.}
\end{figure}

The most stable geometry has an energy of -1.7914 eV. The next energy
found was -1.4982 eV, whose adsorption geometry is depicted in Fig
7.

\begin{figure}
\begin{centering}
\includegraphics[width=12cm]{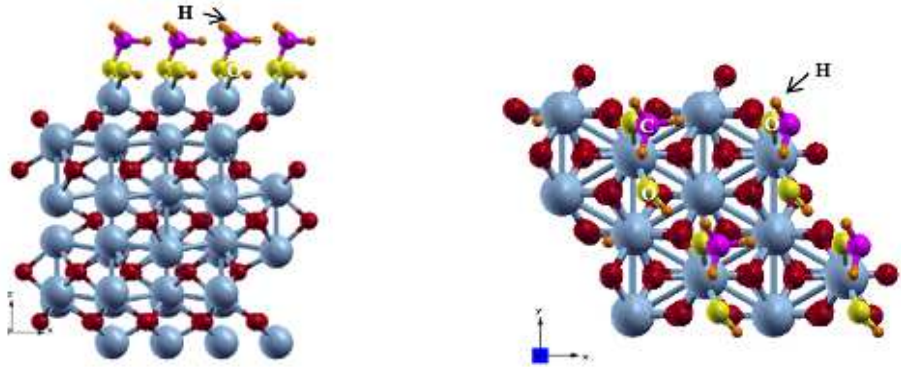} 
\par\end{centering}
\caption{Adsorption geometry ($E_{a}=-1.4982eV$) of $CH_{4}$ on $O_{2}$
dissociatively adsorbed on the $Cr_{2}O_{3}(0001)$ surface.}
\end{figure}

- $CH_{4}$ on $SO_{2}$, preadsorbed on $Cr_{2}O_{3}(0001)$: in
the most stable geometry, with an energy of -0.2758 eV, a hydroxyl
with an $O$ atom of the third substrate layer is formed, while sulfur
binds to the methyl group ($CH_{3}$) and $SO_{2}$ oxygen atoms to
the surface $Cr$ atoms, as shown in Fig. 8.

\begin{figure}
\begin{centering}
\includegraphics[width=12cm]{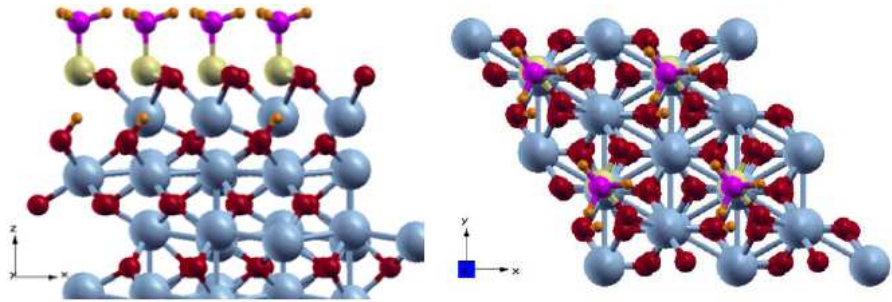} 
\par\end{centering}
\caption{Most stable optimized geometry of $CH_{4}$ on $SO_{2}$ preadsorbed
on the $Cr_{2}O_{3}(0001)$ surface.}
\end{figure}

- $O_{2}$ on $SO_{2}$, preadsorbed on $Cr_{2}O_{3}(0001)$: an adsorption
energy of -1.00 eV was found for the most stable configuration; oxygen
dissociation led to the formation of $SO_{3}$ species on the surface.

\begin{figure}
\begin{centering}
\includegraphics[width=12cm]{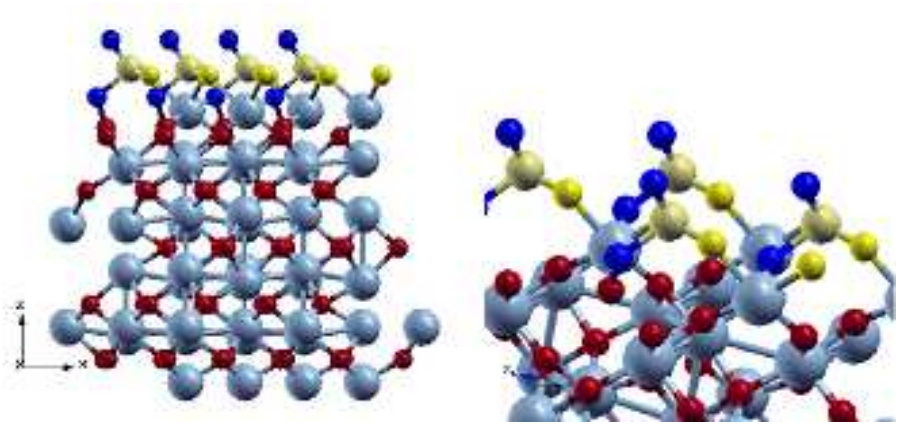} 
\par\end{centering}
\caption{Most stable optimized geometry of $O_{2}$ on $SO_{2}$ preadsorbed
on the $Cr_{2}O_{3}(0001)$ surface.}
\end{figure}

\section{Discussion and Conclusions}

The adsorption of $CH_{4}$ on $Cr_{2}O_{3}(0001)$ is not stable;
the presence of molecularly adsorbed oxygen does not favor adsorption.
However, in the presence of sulfur dioxide or atomic oxygen, $CH_{4}$
may decompose to form hydroxyl species, methyl groups adsorbed on
sulfur and/or methoxy group ($R-O-CH_{3}$) atoms. $SO_{2}$ adsorption
leads to the formation of sulfite species involving surface or preadsorbed
oxygen atoms. No sulfate species formation was observed. $O_{2}$
adsorption in the presence of $SO_{2}$ favors oxygen dissociation
to form sulfite species. Molecular and dissociative $O_{2}$ adsorption
occurs on this substrate with almost the same energy. The most stable
configuration is obtained with the two atoms on the same $Cr$ atom.
The presence of $SO_{2}$ favors the dissociative adsorption of oxygen
to form sulfite species. In previous work \cite{key-4} we experimentally
showed that the activation energy of reaction (3) is lower than that
of reaction (2) under stoichiometric conditions. The presence of $SO_{2}$
would activate the catalyst surface favoring oxygen dissociation and
methane decomposition. FTIR spectroscopy studies on samples of the
catalysts used in reactions (2) and (3) allowed us to identify hydroxyl
and methyl species as the ones found in this work, as well as additional
products of methane dehydrogenations \cite{key-5}. These results
will be presented in a future paper together with further DFT calculations.
The stability of the oxygen molecule adsorbed on the $Cr_{2}O_{3}$
surface may affect the interaction of gas-phase $SO_{2}$ with adsorbed
oxygen and deserves a more detailed study to be included in a future
publication. The size of the cell used in the calculations as well
as periodic boundary conditions simulate experimental conditions with
high adsorbate coverage. The experiments were also performed in a
continuous-flow fixed-bed reactor under conditions that are not directly
comparable to theoretical conditions. Nevertheless, from previous
comparisons with $SO_{2}$ desorption, reliable conclusions from calculations
based on DFT+U can be drawn \cite{key-3}. Then, we conclude that
these calculations may provide useful information on the elementary
stages of reactions (1)\textendash (3) in order to establish the reaction
mechanism.

\section{Acknowledgments }

This work was supported by Consejo de Investigaciones Científicas
y Técnicas (CONICET), Agencia Nacional de Promoción Científica y Tecnológica
(ANPCyT), Universidad Nacional de La Plata and the Faculty of Chemistry
and Engineering Fray Roger Bacon 

\newpage{} 
\end{document}